# Stability Analysis of the Laser System for the TTF Photoinjector at Fermilab

Xi Yang

*Fermi National Accelerator Laboratory*

Box 500, Batavia IL 60510

**Abstract**

A solid-state laser system that produces a 1MHz pulse train of 800 pulses with 18 µJ per pulse at λ = 263.5 nm has been developed to meet the requirements of the TESLA Test Facility (TTF) at Fermilab and in operation since 1998.[1,2]  Besides the production of high charges, high brightness electron beams, the need for high bunch charge stability requires that each laser pulse in the pulse train must have the same energy, and the energy per laser pulse should not vary significantly from shot to shot.  This motivates the stability analysis of the laser system for the TTF photoinjector.

**Laser System**

A mode-locked Nd:YLF oscillator produces a low energy, continuous pulse train at 81.25 MHz.  Pulses from the oscillator are stretched and chirped in a 2 km fiber.  After the fiber, a train of 800 pulses is selected at 1 MHz from the oscillator pulse train by a high speed, low voltage, lithium tantalate (LTA) Pockels cell (Conoptic Inc., model 360-80 modulator).  Each of the 800 pulses are injected into a multipass amplifier which contains a flashlamp-pumped, 1/4×6 inch Nd:glass rod amplifier and a fast KD$^*$P Q-switching Pockels cell (Conoptics Inc., model 350-105).  Each pulse is trapped in the cavity by the Pockels cell and makes 22 passes through the cavity, amplifying up to 40 µJ before it is ejected by the Pockels cell.  A Faraday isolator separates the input and output pulses.  Two flashlamp-pumped, Nd:glass rod amplifier is used in a two-pass configuration to provide an additional gain of 5.  The energy in each pulse after the two-pass amplifier is expected to 200 µJ.  Following the two-pass amplifier, the beam is spatially filtered, and then compressed in time using a pair of parallel diffraction gratings. The pulse length is adjustable in the range of 3 ps to 30 ps. The pulses are frequency doubled and quadrupled in a pair of BBO crystals. The UV pulse train is expanded and transported 20 m in vacuum to a final set of imaging optics, which relays the beam to the photocathode.

The primary challenge in the laser system is in producing uniform amplification for all pulses in the pulse train and pulse-to-pulse stability. The gain of each amplifier can vary substantially during the macropulse (pulse train) depending on the pump rate and the rate at which the stored energy is extracted by amplifier action and spontaneous emission.

Each of the laser amplifiers is driven by a custom designed power supply, which provides nearly constant current discharge to the flashlamps for more than 800 μs. Because the amplifiers are not run to saturation, the amplitude of the output is sensitive to small changes in the gain on shot-to-shot basis. It is preferable to run the multipass amplifier with higher gain and fewer passes to reduce amplitude fluctuations.

### Simplified Model for Stability Analysis

The major contributions to pulse-to-pulse fluctuation of the infrared (IR) pulses before the doubling crystals can be expressed by equation (1) and (2):

$$(I_{out})_n = (G_{mp})^{2 \times N} \times (G_{2p})^2 \times (I_{pp})_n \tag{1}$$

$$\frac{\Delta(I_{out})_n}{(I_{out})_n} = 2 \times \left(\frac{\Delta(G_{2p})}{G_{2p}}\right) + 2 \times N \times \left(\frac{\Delta(G_{mp})}{G_{mp}}\right) + \left(\frac{\Delta(I_{pp})_n}{(I_{pp})_n}\right) \tag{2}$$

Here, $(I_{pp})_n$ is the n$^{th}$ selected pulse from the oscillator, and $(I_{out})_n$ is its amplified pulse after two-pass amplifier, $G_{mp}$ is the gain of multipass amplifier, $G_{2p}$ is the gain of two-pass amplifier. N is the round-trip number of multipass amplifier.

Eq. (2) is obtained by differentiating eq. (1) and we assume that $(I_{out})_n$ is only dependent upon the selected pulse $I_{pp}$, the gain of multipass amplifier $G_{mp}$, and the gain of the two-pass amplifier $G_{2p}$. It is clear from eq. (2) that the contributions to the energy fluctuation of the IR pulses after two-pass amplifier are coming from the gain fluctuation of multipass amplifier, the gain fluctuation of two-pass amplifier, and the energy fluctuation of the selected pulses. Besides that, the contribution from multipass amplifier is N times larger than that of two-pass amplifier whenever the gain of multipass amplifier fluctuates at the same level with that of two-pass amplifier.

## Experimental Results

One of the major contributions to the gain fluctuation of the flashlamp-pumped amplifier is the capacitor bank voltage fluctuation. The flashlamp is fired when the capacitor bank of the power supply is discharging. Since there is a square relationship between the flashlight intensity and the capacitor bank voltage, for example, a 2% increase of the capacitor bank voltage will bring a 4% increase of the flashlight intensity, and this will cause a 4% increase of the gain. It often takes eleven round trips for a selected pulse to be amplified by multipass amplifier, and according to eq. (2), there is a 137% (*$1.04^{11\times2}-1$*) energy increase after multipass amplifier. Therefore, we did the following experiment: we measured the energy of amplified pulses after multipass amplifier as a function of the capacitor bank voltage of the flashlamp power supply. Each data point was taken by a 120-shot average of the laser energy meter in order to minimize the influence of the selected pulse fluctuation. The result is shown in Fig. 1, and it is consistent with our analysis.

Secondly, we did the measurement about the correlation between the selected pulse from the oscillator and its amplified pulse after multipass amplifier. A schematic of the experimental setup is shown in Fig. 2. We used cable delay to match the selected pulse and its amplified pulse after multipass amplifier and connected them to a Lecroy *540* scope in order to use x-y plot to find their correlation. In Fig. 3, x-axis represents the selected pulse, and y-axis represents the amplified pulse. Line AB, which is nearly perpendicular to x-axis, indicates the situation that the amplified pulses after multipass amplifier can vary in a wide range even if their selected pulses have the same intensity. Here, the intensity fluctuation of the amplified pulses after multipass amplifier can be expressed by the ratio between $y_B$ and $y_A$, which is 2.6, and from which we can estimate that the gain fluctuation of multipass amplifier is about 160%. We also did the correlation measurement on the pulses before and after two-pass amplifier. The result is shown in Fig. 4. Since all the points are located in the region between line 1 and line 2 with a slope of 19.5 and 23 separately, the maximum of the gain fluctuation of two-pass amplifier is about 16%.

It is clear from the above experiments that the major contribution to the energy fluctuation of the IR pulses before doubling crystal is the gain fluctuation of mulitpass amplifier. The cause of the gain fluctuation of flashlamp-pumped amplifier that can be measured in our lab is the fluctuation of the capacitor bank voltage, which is about 1%. From the above analysis, the 1% fluctuation in the capacitor bank voltage brings about 55% intensity fluctuation of the multipass amplifier output. Based upon the data shown in Fig. 3, the intensity fluctuation of the multipass amplifier is greater than 100%. There are some other sources, which contribute to the gain fluctuation of multipass amplifier, such as the room temperature fluctuation, etc. Fig. 5 shows the output of mulitpass amplifier at a constant selected pulse rate in one hour, and there is an intensity fluctuation at a period of about 12 minutes. This is consistent with the room temperature change in the laser room.

**Conclusion**

It is clear that the existed laser system for TTF photoinjector at Fermilab needs to be improved in order to satisfy the requirements of producing high charge, high brightness and bunch-to-bunch stable electron beams. The gain of multipass amplifier, which is higher than 6000, greatly reduces the number of two-pass amplifiers and simplifies the laser system at Fermilab. The drawback of multipass amplifier is that its gain fluctuation brings an exponential increase in the intensity fluctuation of the output pulses. Fast feedback control system needs to be installed to flashlamp-pumped amplifiers, especially multipass amplifier in the future in order to obtain the uniformity for all pulses in the 800 μs pulse train and train-to-train stability.

Reference:
[1] Alan R. Fry, "Novel Pulse Train Glass Laser for RF Photoinjectors" PhD thesis, University of Rochestor, 1996.
[2] A. R. Fry, M. J. Fitch, A. C. Melissinos, and B. D. Taylor, "Laser system for a high duty cycle photoinjector" *Nuclear Instruments and Methods in Physics Research A*, 430:180-188, 1999.

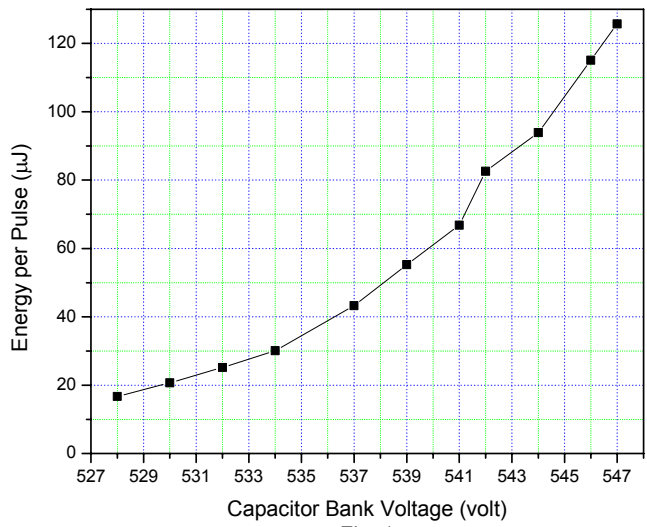

Fig. 1 The energy of an amplified pulse after multipass amplifier *vs*. the capacitor bank voltage of the flashlamp power supply.

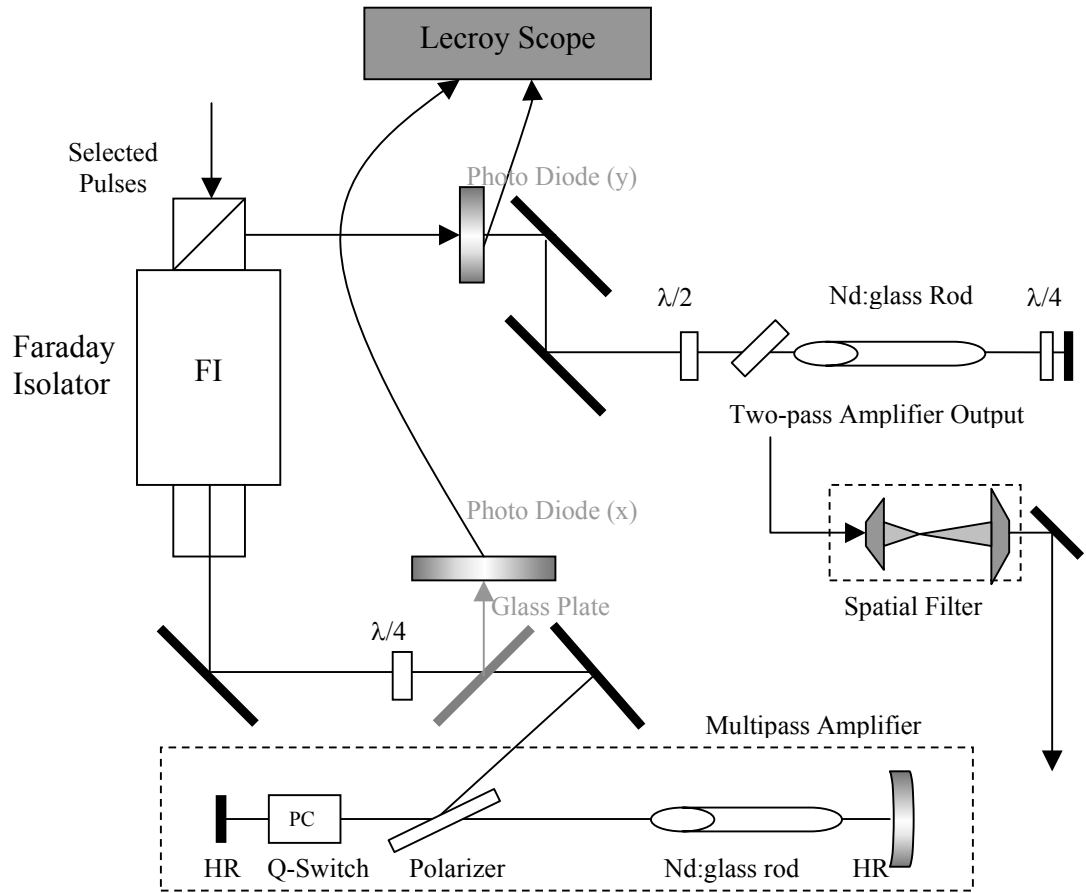

Fig. 2

Fig. 2 Schematic of the experimental setup.

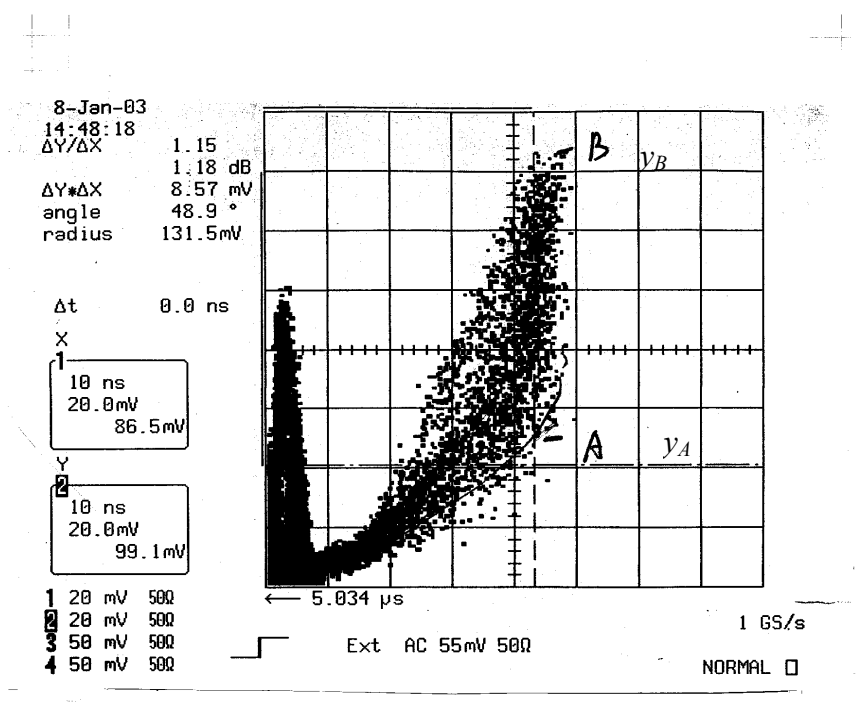

Fig. 3

Fig. 3 x-y plot. x-axis represents the selected pulses, and y-axis represents the amplified pulses after multipass amplifier.

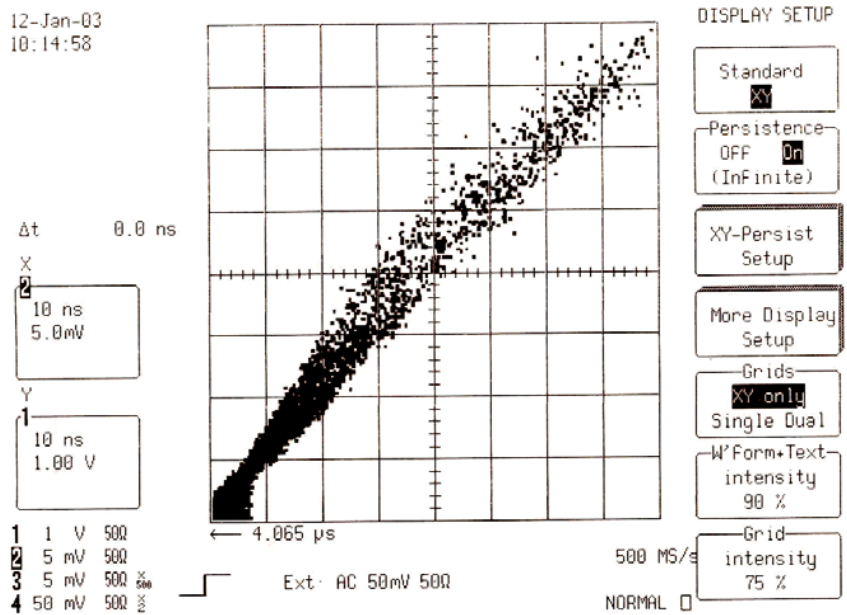

Fig. 4

Fig. 4 x-y plot.  x-axis and y axis represent the pulses before and after two-pass amplifier separately.

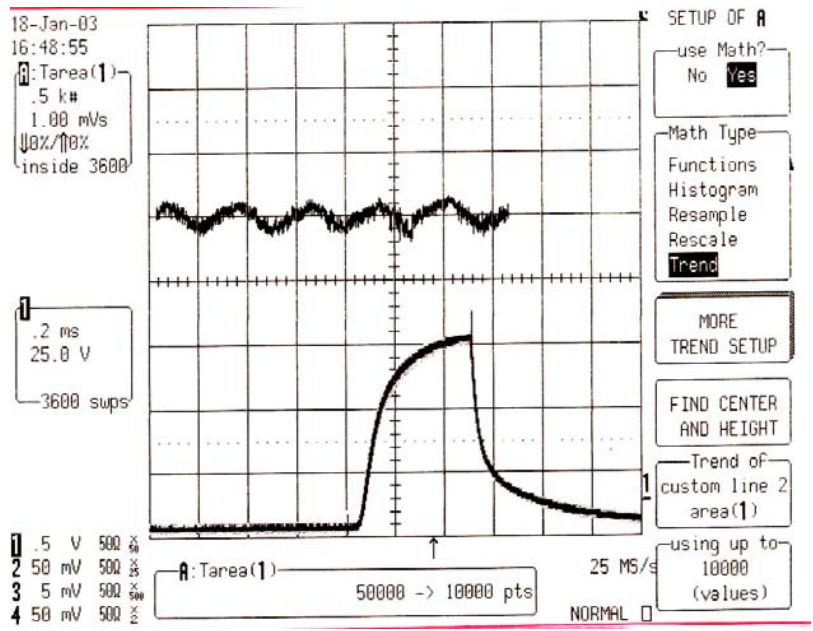

Fig. 5

Fig. 5 The top trace is the output of mulitpass amplifier at a constant selected pulse rate in one hour, and it includes five periods of the multipass output fluctuation.